\documentstyle[twocolumn,prl,aps,psfig,floats]{revtex}
\begin{document}
\draft
\narrowtext
\title{\Large\bf Unusually Large
Thermal Expansion of Ag(111) }
\author{Shobhana Narasimhan and Matthias Scheffler}
\address{
Fritz-Haber-Institut der Max-Planck-Gesellschaft,\\ Faradayweg 4-6,
D-14195 Berlin-Dahlem, Germany}

\date{\today}
\twocolumn[
\maketitle
\begin{quote}
\parbox{16cm}{\small

We investigate the thermal behavior of the (111) surface of silver, using
phonon
frequencies
obtained from {\it ab initio} total energy calculations, and anharmonic effects
 treated
within a quasiharmonic approximation.
Our results reproduce the experimental
observation of a large and anomalous increase in the surface thermal expansion
at high
temperatures\cite{statiris94}. Surprisingly, we find
that this increase can be attributed to a rapid softening
of the {\it in-plane}
phonon frequencies, rather than due to the anharmonicity of the out-of-plane
surface phonon modes. This
provides evidence for a new mechanism for the enhancement of surface
anharmonicity. A comparison with Al(111)
shows that the two surfaces behave quite differently, with no evidence for
such anomalous behavior on Al(111).
}
\end{quote}]

The equilibrium lattice constant of a crystal is determined by
a balance between the various attractive and repulsive forces
present within the solid. When the crystal is cleaved to form surfaces, this
balance is destroyed, and the atoms at the surface therefore relax
either
inwards or outwards. Both calculations and experiments show that
the low-index surfaces of most metals, including the noble metals, relax
inwards (at temperature $T \rightarrow 0$ K),
in accordance with simple bond strength - coordination number arguments.

However, recent experiments \cite{statiris94} show that
this situation is dramatically reversed for the Ag(111) surface as the crystal
is heated:
Up to
$T \approx 670$ K, the first interlayer spacing $d_{12}$ is indeed
{\it contracted} by $\sim$ 2.5\% relative to the bulk separation
$d_B$; but upon increasing $T$ further, $d_{12}$ increases much more rapidly
than
$d_B$ does, so that by 1150 K it is {\it expanded} by $\sim$ 10\%.
Correspondingly,
the surface coefficient of thermal expansion $\alpha_S$ becomes more than ten
times as large as the bulk coefficient of thermal expansion
$\alpha_B$ \cite{explain}. Since thermal
expansion arises from the anharmonicity of the interatomic potentials, this in
turn
indicates a significant enhancement in anharmonicity at the surface.

It has long been realized that it is possible for measures of anharmonicity
(such as coefficients of thermal expansion,
mean squared displacements of atoms,  and the rate of change of
phonon frequency with temperature) to be larger at surfaces than in bulk
crystals\cite{allen69};
and the conventional lore recognizes two possible
reasons for such an enhancement: {\it (i)} the breaking of symmetry due to the
presence of
the surface makes the interlayer potential more asymmetric at
the surface than in the bulk, increasing the size of the
odd terms in
a Taylor series expansion
of the energy in powers of atomic displacements;
{\it (ii)} atomic displacements may be
greater at the surface than in the bulk, thus increasing the
relative magnitudes of the higher-order anharmonic terms in this series
expansion.
However, early experiments and calculations
\cite{wilson71},\cite{dob73ma78} on metal
surfaces showed that  measures of anharmonicity
are
typically one to three times larger at the surface than in the bulk, and thus
the
significantly
larger enhancement seen on Ag(111) is remarkable, and is especially unexpected
for a close-packed face-centered-cubic (fcc) (111) surface.

Moreover, there is a serious disagreement between experiments and calculations
on the Ag(111) surface:
Lewis\cite{lewis95} has carried out molecular-dynamics simulations using
interatomic
potentials obtained using the embedded-atom-method (EAM) to investigate the
thermal behavior of Ag(111). The
results of these simulations differ significantly from those reported
experimentally: The surface layer relaxes inwards at all temperatures,
and $\alpha_S$ is less than twice as large as $\alpha_B$.

We note that there have been other recent experimental observations of large
enhancements in
surface anharmonicity, but on more open surfaces, {\it viz.}
Pb(110)\cite{frenken87},
Ni(001)\cite{cao90} and Cu(110)\cite{helgesen93}. In contrast to the situation
for Ag(111),
the results of simulations using EAM potentials  to
study Cu(110)\cite{yang91} and Ni(001)\cite{beaudet93}
show the same qualitative features as the experimental data,
though the calculated enhancement is somewhat smaller than
that measured experimentally.

Why then is there a disagreement between experiment and calculation for
Ag(111)? Is it due to
inadequacies of the EAM potentials? Or could it be an indication of some
hitherto
undetected surface phase transition?

To study these questions,
we have investigated the anharmonic properties of Ag(111) and
(for purposes of comparison) Al(111)
by performing {\it ab initio} calculations using density functional theory.
Fully separable norm-conserving pseudopotentials\cite{gonze91}
were used in our calculations, together with a plane wave
basis set with an energy cut-off of 60 Ry (20 Ry for Al), and  the
local-density
approximation with Ceperley-Alder exchange-correlation\cite{ceperley80}.
We have verified that these pseudopotentials provide a satisfactory
description of harmonic as well as
anharmonic properties of bulk Ag and Al \cite{verification}.

The surface calculations were performed using a repeated slab geometry
consisting of six
atomic layers separated by a vacuum layer of the same thickness.
The {\bf k}-point sets used to sample reciprocal space contained seven points
in the irreducible part of the Brillouin zone for the undistorted  surface;
the number of {\bf k}-points was correspondingly increased upon breaking
symmetries
by distorting the lattice in order to calculate phonon frequencies.
Convergence
of calculated anharmonic quantities
with respect to energy cut-off, number of {\bf k}-points and number of layers
was carefully
tested for.

Our strategy is to compute static energies and phonon frequencies by performing
self-consistent calculations at $T=0$ K, and then extend our results to finite
temperatures
by using a quasiharmonic approximation.

In order to obtain a qualitative understanding of the mechanisms in operation,
and a first estimate
of the size of surface anharmonic effects,
we use a simple model
of the lattice dynamics of the surface, considering only three phonon modes, in
 all of
which the topmost layer of the slab moves as a whole -- i.e., we assume that
the
displacements are confined to the first layer of atoms at the surface, and
consider
only those modes with zero wave-vector. We consider one mode in which the
surface
atoms vibrate normal to the surface plane (along the $z$ direction), and two
modes
(along ${\bf x}$ = $[1 { \overline1} 0]$ and ${\bf y}$ = $[1 1 {\overline2}]$)
in which they vibrate in the
plane of the surface.

We first computed the change in the total
energy of the Ag(111) slab upon varying the first two interlayer separations
$d_{12}$ and
$d_{23}$, and maintaining the fcc stacking of the bulk crystal.
This not only provides the static interlayer potential, but is also equivalent
to simulating the vibrational mode along $z$. Our result for the dependence on
$d_{12}$ of the
first interlayer potential is plotted in Fig.~1; it is clearly asymmetric about
the minimum at $d_{12} = 2.30$ \AA. We found that
allowing for the relaxation of  $d_{23}$ does not have a significant impact on
the
results for the close-packed (111) surface\cite{d23}.
 For
the results presented in Fig.~1 and the rest of this paper, $d_{23}$ is
therefore fixed at the bulk
interlayer separation of $2.34$ \AA.

To see how this anharmonicity of the interlayer potential is manifested at
finite
temperatures, we consider a one-dimensional quantum oscillator vibrating in the
 interlayer
potential of Fig.~1. A numerical solution of the Schr\"odinger equation for
this problem furnishes
the eigenstates and eigenvalues of such an oscillator, and  the mean
displacement
$\langle d_{12}\rangle_n$  in the $n$-th eigenstate is obtained by computing
the
expectation value of the displacement operator in each state. The average value
 at a
finite temperature $T$ is then obtained by weighting these results with the
corresponding
partition function.

Our results for
$d_{12}(T)$ obtained from this procedure (see the open
circles in Fig.~3) indicate a
modest enhancement in $\alpha_S$ relative to $\alpha_B$ of $\sim 1.7$, which is
 in
agreement with the EAM results\cite{lewis95}, but much
smaller than that measured experimentally.

\begin{figure}[b]
{{
\hspace{0.5cm}
\psfig{figure=  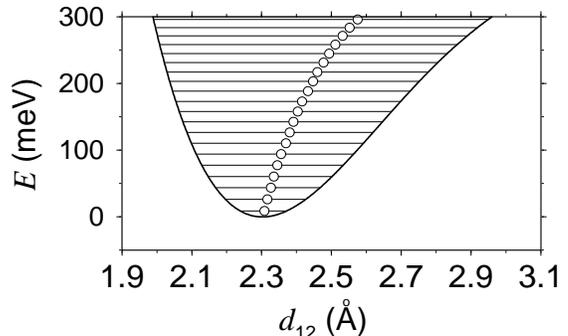 ,width=7.5cm}
}}
\caption{Static interlayer potential between the two outermost layers
of Ag(111). The horizontal lines and open circles indicate the
energy eigenvalues $E_n$ and the mean displacements
$\langle d_{12}\rangle_n$ respectively of an Ag atom vibrating in this
potential. }
\end{figure}

\begin{figure}[b]
%
{{
\hspace{0.5cm}
\psfig{figure=  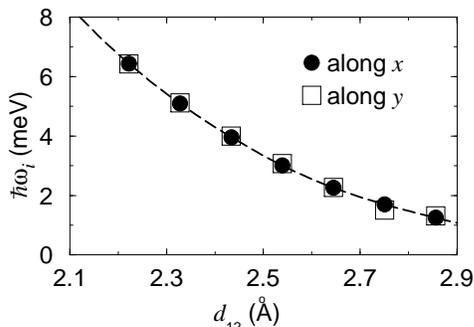 ,width=7cm}
}}
\caption{Energy $\hbar\omega_i$ of the top-layer in-plane modes of
Ag(111), as a function of the first interlayer separation $d_{12}$. Circles
and squares indicate modes polarized along the $x$- and $y$- directions
respectively.}
\end{figure}

We next performed ``frozen-phonon" calculations
to study the behavior of the two in-plane modes
in our model. At each value of  $d_{12}$, the atoms in the surface layer were
displaced along
first the $x$ and then the $y$ direction, and the total energy was computed for
 a series of
displacements up to $\pm 0.15$ \AA. The curvature of the resulting plots of
energy versus
displacement gives the mode frequency.
We find that the frequency of these in-plane modes
decreases surprisingly rapidly upon increasing $d_{12}$; these results are
plotted in Fig.~2. The value of the Gr\"uneisen parameter $\gamma$ extracted
from our results for the in-plane modes (between
5.5 and 7.5) is significantly larger than the average  $\gamma$ of 2.46 for
bulk
modes\cite{gschneider64}.
Thus, the frequently made assumption\cite{wilson71} that $\gamma$ is
approximately equal for surface and bulk modes
is clearly invalid in this case.

The surface can thus reduce its vibrational free energy significantly by
expanding outwards, though
such
an outward expansion would be accompanied by an increase in the static
energy. The optimal
value of $d_{12}$ is determined by minimizing the free energy\cite{allen69}:

\begin{equation}\label{eq:ftot}
F(d_{12},T)  = E_{\rm stat}(d_{12}) + \sum_{i=x,y,z}
F_{\rm vib}^{i}(d_{12},T) ,
\end{equation}
\noindent
at each temperature $T$ . Here, $E_{\rm stat}(d_{12})$ is the static
interlayer potential plotted in
Fig.~1, and $F_{\rm vib}^i(d_{12},T)$ is the vibrational free energy
corresponding to vibrations in
the $i$-th direction, which, in the quasiharmonic approximation, is given
by\cite{allen69}:
\begin{equation}\label{eq:fvibqh}
F_{\rm vib}^{i}(d_{12},T) = k_B T ln \Bigl\{2 sinh \Bigl({{\hbar\omega_i(d_{12}
)}\over{2 k_B T}}\Bigr)\Bigr\};
\end{equation}
\noindent
where $k_B$ and $\hbar$ are Boltzmann's constant and Planck's constant
respectively.
The frequency of the mode polarized along the $i$-th direction, when the first
interlayer
separation is fixed at $d_{12}$, is denoted by $\omega_i(d_{12})$; its value
is obtained directly from our frozen-phonon calculations for the
two in-plane modes (see Fig.~2). For the out-of-plane mode, we compute
$F_{\rm vib}^z(d_{12},T)$ numerically, in such a
way as to reproduce the exact result for $d_{12}(T)$ that we have already
obtained by solving the
Schr\"odinger equation when only the mode along $z$ is present.

\begin{figure}[t]
\par
\par
{{
\psfig{figure=  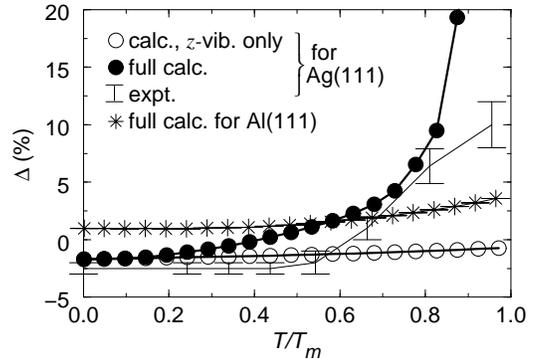 ,width=8.5cm}
}}
\caption{Contraction/expansion of $d_{12}$ relative to $d_B$, as a function of
normalized temperature $T/T_m$ --
our calculations for Ag(111) with out-of-plane mode only (open circles),
and all three modes (filled circles); experiment$^1$ on
Ag(111) (solid line with error bars); our calculation for Al(111)
with all three modes (stars). $T_m$ is the bulk melting temperature
($T_m^{Ag}$ =
1234 K, $T_m^{Al}$ = 933 K).}
\end{figure}

Our final result for $d_{12}(T)$ in the presence of all three modes, obtained
by minimizing the
free energy given by Eq.~(1), is given by the filled circles in Fig.~3 (note
that the temperatures
are normalized with respect to the melting temperature $T_m$).
The experimentally measured data points\cite{statiris94} are also plotted;
both the
experimental and theoretical curves display the same features:
there is little or no change up to about $T/T_m = 0.5$, i.e., $T = 617$ K,
beyond which there is a rapidly increasing trend towards
outwards relaxation of the surface layer.
At $T/T_m = 0.85$, i.e., $T = 1049$ K, we find that the surface layer is
relaxed outwards
by about 15\%, whereas
the experiments show an outwards relaxation of $\sim 7.5\%$. Given the
simplicity of our model,
and the large experimental error bars, this is as good an agreement as we can
hope for.

The increasing slope of $d_{12}(T)$ reflects a flattening
in the minimum of the free-energy curve, and the rapid increase in $d_{12}(T)$
at high $T$ is a precursor to the development of a saddle-point
instability in the free-energy curve, similar to that which has
been obtained in studies of the surface melting of copper surfaces
\cite{jayanthi}.  We emphasize that the rapid decrease of $\omega_i(d_{12})$
for the in-plane
modes is crucial for obtaining the large outwards expansion; if, for example,
this rate of decrease
were to be halved,
the maximum outwards expansion would be  drastically reduced to
about 2\%.

We expect that our numerical results will change slightly upon including other
surface phonon modes and
allowing for dispersion through the surface Brillouin zone; however, we note
that measurements of surface phonon frequency-shifts suggest that
the degree of anharmonicity remains approximately constant through a
surface phonon band\cite{benedek92}. We should also allow
for expansion in the plane of the surface, but this should in fact reinforce
the
softening of the in-plane modes.

\begin{figure}[t]
{{
\hspace{1cm}
\psfig{figure=  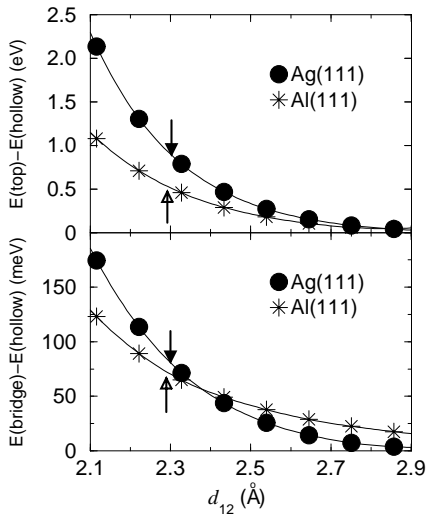 ,height=7cm}
}}
\caption{Decrease in the corrugation of the interlayer potential with
increasing interlayer separation $d_{12}$: the  graphs show
the increase in surface energy when the outermost layer of atoms
occupies an atop site or bridge site instead of the favored fcc hollow site,
for Ag(111) (filled circles) and Al(111) (stars). Arrows indicate the
equilibrium
value (neglecting zero-point vibrations) of $d_{12}$ at $T$ = 0 K.}
\end{figure}

To check whether such behavior is universal or a peculiar property of Ag(111),
we repeated the same calculations
on bulk aluminum and Al(111). Our results for the relaxation of $d_{12}$ for
Al(111) are also plotted in
Fig.~3, and it is obvious that there is no evidence for a dramatically
increased surface anharmonicity
on Al(111).

Our results indicate that in addition to the two well known  sources of
enhanced surface
anharmonicity that we have already mentioned, a third
(and hitherto neglected)
effect is more important in causing
the dramatic enhancement in surface
anharmonicity on Ag(111): Not only do interlayer potentials at the surface
tail off
rapidly with increasing $z$ , they
simultaneously {\it become much flatter in the xy plane} -- in other
words, the operative effect is not so much a decrease in the absolute
magnitudes
of interlayer
potentials at the surface, but a {\it reduction in their corrugation parallel
to
surface}. As a consequence, those surface phonon modes in which atomic
dispacements have significant components in the surface plane soften rapidly
upon increasing interlayer separations; this drives the outermost layer of
atoms to
expand outwards at high temperatures.

An
accurate description of the in-plane corrugation clearly requires taking into
account the
correct distribution of the electronic charge density (and the resulting
chemical bond
formation) at the surface. It seems plausible that when rebonding effects
become
significant, the EAM (which essentially ignores the relaxation of atomic
charge densities
and the rehybridization of electronic states) may fail; this may be why
Lewis \cite{lewis95}
did not observe
a large outwards expansion in his  simulations.

The rapid decrease in the corrugation of the interlayer potential is evident
in Fig.~4,
where we have plotted the differences
in energies when the outermost layer of atoms occupies various stacking sites.
Note that
{\it (i)} the flattening occurs more rapidly for Ag(111) than Al(111) {\it
(ii)} upon allowing for
the lighter mass of Al atoms, the {\it effective} corrugation relevant for
phonon
frequencies is actually larger for Al(111) than for Ag(111).  Both these
factors contribute
to the enhancement in $\gamma$ of the in-plane top-layer modes and thus the
larger thermal expansion of Ag(111).

Such a decrease in the corrugation of the substrate potential would also tend
to favor a top-layer
reconstruction of the type that has been observed on Au(111)\cite{au111expt}
or Pt(111)\cite{pt111expt},
where the substrate potential is too weak to prevent a densification of atoms
in the topmost layer.
However,  experiments
apparently show no evidence of such a reconstruction on
Ag(111)\cite{statiris94}; further calculations of surface stresses and the
strength of intralayer couplings should help clarify the situation.

Our conclusion that the enhancement in surface anharmonicity arises mainly
from in-plane vibrations is supported
by the experimental observation that the magnitude of in-plane vibrations on
Ag(111) rises faster than the
magnitude of out-of-plane vibrations\cite{gustpriv}.
Experiments on other surfaces, e.g., Cu(001), have detected in-plane
vibrational
amplitudes that are larger than out-of-plane amplitudes\cite{jiang91},
and it is interesting to speculate whether this
(counter-intuitive) result arises from the same cause, i.e., from a rapid
softening of in-plane frequencies due to
thermal expansion.

In conclusion, we have demonstrated that the anomalously large surface thermal
expansion
of Ag(111) can be attributed to a rapid softening of in-plane vibrational
modes (related
to a rapid flattening of the corrugation of the interlayer potential) upon
increasing
interlayer distances.
We have shown that a similar scenario does not, however, lead to significant
enhancement on Al(111).



\begin{references}
\bibitem{explain}
Throughout this paper, when we refer to ``surface thermal expansion", it is
implied
that we are referring to changes in interlayer separations. It is assumed that
intralayer
separations are the same in the bulk and at the surface.
\bibitem{verification}
 e.g., the lattice constant $a_0$ of Ag(Al) is 4.06
(3.94) \AA, the bulk modulus is  1.22 (0.80) MBar, and the pressure
derivative of
the bulk modulus is 5.94 (4.22); all of these values compare well with
experiment.
\end{references}
\end{document}